# Withholding or withdrawing invasive interventions may not accelerate time to death among dying ICU patients

Daniele Ramazzotti, PhD[1], Peter Clardy, MD[2], Leo Anthony Celi, MPH[3], David J. Stone, MD[4], Robert S. Rudin, PhD[5]

[1]Department of Pathology, Stanford University, Stanford, CA, United States
[2]Departments of Medicine and Medical Education, Division of Pulmonary and Critical Care, Mount Auburn Hospital, Cambridge, MA and Harvard Medical School, Boston, MA, United States
[3]Institute for Medical Engineering and Science, Massachusetts Institute of Technology, Cambridge, MA, United States
[4]Departments of Anesthesiology and Neurological Surgery, University of Virginia School of Medicine, Charlottesville , VA, United States
[5]RAND Corporation, Boston, MA, United States

**Corresponding author:**
Leo Anthony Celi,
Institute for Medical Engineering and Science, Massachusetts Institute of Technology,
Cambridge, MA, United States
lceli@mit.edu

**Keywords**

Terminal care, Intensive care, Intensive care units, Trends, Survival

**Abstract**

**Background:** Critically ill patients may die despite invasive intervention. In this study, we examine trends in the application of two such treatments over a decade, namely, endotracheal ventilation and vasopressors and inotropes administration, as well as the impact of these trends on survival durations in patients who die within a month of ICU admission.

**Methods:** We considered observational data available from the MIMIC-III open-access ICU database and collected within a study period between year 2002 up to 2011. If a patient had multiple admissions to the ICU during the 30 days before death, only the first stay was analyzed, leading to a final set of 6,436 unique ICU admissions during the study period. We tested two hypotheses: (i) administration of invasive intervention during the ICU stay immediately preceding end-of-life would decrease over the study time period and (ii) time-to-death from ICU admission would also decrease, due to the decrease in invasive intervention administration. To investigate the latter hypothesis, we performed a subgroups analysis by considering patients with lowest and highest severity. To do so, we stratified the patients based on their SAPS I scores, and we considered patients within the first and the third tertiles of the score. We then assessed differences in trends within these groups between years 2002-05 vs. 2008-11.

**Results:** Comparing the period 2002-2005 vs. 2008-2011, we found a reduction in endotracheal ventilation among patients who died within 30 days of ICU admission (120.8 vs. 68.5 hours for the lowest severity patients, $p<0.001$; 47.7 vs. 46.0 hours for the highest severity patients, $p=0.004$). This is explained in part by an increase in the use of non-invasive ventilation. Comparing the period 2002-2005 vs. 2008-2011, we found a reduction in the use of vasopressors and inotropes among patients with the lowest severity who died within 30 days of ICU admission (41.8 vs. 36.2 hours, $p<0.001$) but not among those with the highest severity. Despite a reduction in the use of invasive interventions, we did not find a reduction in the time to death between 2002-2005 vs. 2008-2011 (7.8 days vs. 8.2 days for the lowest severity patients, $p=0.32$; 2.1 days vs. 2.0 days for the highest severity patients, $p=0.74$).

**Conclusion:** We found that the reduction in the use of invasive treatments over time in patients with very poor prognosis did not shorten the time-to-death. These findings may be useful for goals of care discussions.

**Background**

Critically ill patients commonly receive invasive interventions including endotracheal ventilation and continuous intravenous such as vasopressors and inotropes. These kinds of treatments may be administered even to patients with very poor prognosis in an attempt to sustain life. About 500,000 adults die each year in the United States during or shortly after ICU admission[1]. Most of these deaths occur after decisions are made to transition the goal of treatment from cure to comfort measures[2,3]. Healthcare proxies often make these decisions because the decision-making capacity of the patient is impaired[4,5].

However, invasive treatments are often supportive rather than curative. Many if not most of patients so identified will not survive beyond a relatively short term, and are among the more than 1 in 5 patients in the US who die after receiving what proved to be "non-beneficial" care in the intensive care unit (ICU)[6]. Endotracheal ventilation at the end of life deprives patients of the ability to speak with family and friends, which may severely diminish the patient's quality of life in their final moments, and may engender distress among their surviving family and friends. However, little is known about how rates of invasive interventions for end-of-life patients in the ICU have changed over time, or the impact of any such changes on how long these patients ultimately live.

To explore these questions, we examined trends in the application of invasive interventions in the ICU at one institution over the course of ten years. We hypothesized that rates of invasive intervention for patients who died during or soon after an ICU stay would decrease because of the introduction of efforts to improve end-of-life communication and care in the ICU. We also hypothesized that with these invasive treatments withheld or discontinued, these patients would likely die sooner. Importantly, the end of invasive interventions for patients in this study did not indicate the end of medical and nursing care. After discontinuation of invasive measures, comfort and symptom control measures already in place were continued at the end of life.

**Methods**

Clinical data were extracted from the MIMIC-III open-access ICU database[7]. The database comprises of almost 60,000 hospital admissions for a total of 38,645 unique adults, collected from June 2001 to October 2012 in the ICUs of the Beth Israel Deaconess Medical Center, Boston, Massachusetts.

**Cohort of study.** Our analysis included all adult patients (18 years old or older) who died within 30 days of their ICU admission. We considered a study period between year 2002 up to 2011. If the patient had multiple admissions to the ICU during the 30 days before death, only the first stay was analyzed, leading to a final set of 6,436 unique ICU admissions during the study period (see Figure 1).

**Study procedure.** We explored the following hypotheses: (i) administration of invasive intervention during the ICU stay immediately preceding end-of-life would decrease over the study time period and (ii) time-to-death from ICU admission would also decrease due to the decrease in invasive intervention administration. We compared results for the first hypothesis with patients who survived hospital stays. Invasive intervention was defined by the following variables: vasopressors and inotropes administration, which required insertion of a central intravenous access (as both binary variable yes/no and duration) and endotracheal ventilation (as both binary variable yes/no and duration). We also examined utilization trends of red blood cell transfusion, renal replacement therapy, extracorporeal life support, and existence of a "do not resuscitate" (DNR) order. Time-to-death was computed from the time of ICU admission. The following variables were extracted to adjust for confounding: age at admission, gender (binary variable, male or female), Simplified Acute Physiology Score or SAPS I[8], Elixhauser Comorbidity Index score[9] and Do Not Resuscitate (DNR) status at admission. All data extraction queries were performed using PostgreSQL (https://www.postgresql.org). We analyzed trends for the entire cohort of patients, and for subgroups with highest and lowest SAPS I score, to assess if any effect would be more pronounced for the sickest patients.

**Subgroup Analysis.** To further investigate the second hypothesis, we conducted a subgroup analysis to examine among the patients with lowest and highest severiy whether we could identify a subset of patients who actually survived longer when invasive interventions were withheld. Note that 'lowest' severity score is a relative term, as all these patients ultimately died within 30 days of ICU admission, and therefore had absolute severities higher than the general ICU population would manifest. To this extent, we stratified the patients based on their SAPS I scores, and then focused on the patients within the first and the third tertiles. The patients in the first tertile represent the lowest severity score, while the patients in the third tertile represent the highest severity score. We then performed our analysis in these two subsets and specifically, we compared trends for these groups between years 2002-05 vs. 2008-11.

**Statistical Analysis.** The analyses were performed using the open-access software R 3.4 (http://www.R-project.org/). Tests for significant differences were performed for both binary and continuous variables. This was assessed by Chi-square test[10] for binary variables and by t-test test[11] for continuous variables. A p-value threshold of 0.05 was used to assess statistical significance.

**Results**

Characteristics for our cohort of 6,436 patients across ten years are reported in Table 1. Age at admission, gender and SAPS I scores were similar during the study period, showing overall comparable clinical conditions for the patients admitted to the ICUs during those ten years. The single exception was an observed increase for Elixhauser scores, indicating that the patient population had more co-morbidities and likely a higher level of prior-to-admission clinical severity (Table 1). Over the study period, we observe a significantly decreasing trend in the administration of invasive interventions (use of vasopressors and inotropes and endotracheal ventilation, see also Table 2), which was consistent with our first hypothesis. This trend was also seen among the ICU patients who survived their hospitalization (see Figures 2-3). The use of non-invasive ventilation increased (see Table 2) during the study interval. We also found a reduction in the administration of red blood cell transfusion among the ICU patients who died during the study period but observed a slight but not significant up-trend in the use of renal replacement therapy in this cohort (data not shown). The use of extracorporeal life support was steady at <1% among those who died throughout the study period (data not shown).

Despite a reduction in the use of endotracheal ventilation and vasopressors and inotropes over time among ICU patients who did not survive, and contrary to our second hypothesis, we found no change in median time-to-death from admission to the ICU (Table 2). There was an upward trend in the number of patients with a "do not resuscitate" or DNR order at the time of ICU admission (see Table 2). The proportion of patients who died as a result of unsuccessful cardio-pulmonary resuscitation was very low throughout the study period at <1%, suggesting that the vast majority of patients died because of termination of curative treatments (data not shown).

The results for the lowest severity score patients resembled the trends observed in the whole population. However, the highest severity score patients show slightly different results; endotracheal ventilation use followed a similar decreasing trend as the overall group, but the use of vasopressors and inotropes remained constant (i.e., no statistical difference) during the study period (75% vs 71% patients were administered vasopressors and inotropes in the two earliest and most recent years for a median duration of 24 and 27 hours respectively). Time-to-death remained unchanged in both groups. To more clearly present the findings within the subgroups, we show average results of these two groups of patients during a period of the earliest years (2002 to 2005) and the most recent years (2008 to 2011) (Table 2).

**Discussion**

While clinicians cannot predict whether a patient will die before a certain time, they can identify patients with particularly poor prognoses. We included such patients by extracting data from the ICU subpopulation that did not survive more than a month past ICU admission from 2002 to 2011 to derive insight into the impact of invasive interventions on their eventual outcomes. Midway through this time period, an educational curriculum was introduced to the ICU residents that provided training in end-of-life communication skills. Though observational, our analyses of time trends in one large institution suggest that in accordance with our hypothesis, these interventions are being employed less frequently in this group. This trend may be partly explained by an increase in patients with DNR orders, which was likely the effect of an end-of-life communication course that was introduced in 2008 to the residents rotating in the medical ICU, as well as heightened discussion around advanced directives in the mainstream media. However, contrary to our hypothesis, reducing invasive interventions for critically ill patients did not result in a shorter time to death. These results demonstrate that for patients identified to have extremely poor prognoses, longevity is unimpaired when invasive interventions are withheld or withdrawn. In fact, the quality of the remaining life available to these patients is very likely to be enhanced by the absence of such interventions.

To our knowledge, this is the first investigation into trends of invasive intervention for end-of-life care and the impact of their utilization on time to death. However such findings are not without precedent. One study of palliative care patients found that such care can result in reduced use of invasive interventions and higher quality of life in terminal metastatic non-small cell lung cancer patients[12]. Other studies[13] have found reduced lengths of stay in an ICU after a separate palliative care unit was opened. The increase in the use of non-invasive ventilation during the study interval may account for some (or all) of the survival observed in patients in whom endotracheal ventilation was discontinued[14]. It is less clear why the discontinuation of vasopressors and inotropes support should result in little survival change. Perhaps when endotracheal ventilation was not initiated or discontinued, the cardiovascular system may benefit from removal of the adverse effects of positive pressure ventilation on cardiovascular physiology. It is also possible that the clinicians are more likely to accept values outside the "norm" for blood pressure and/or oxygen saturation than they would for patients with better expected outcomes.

The trend we observed toward decreasing use of invasive intereventions at the end of life has been previously reported. One recent study within a French ICU found a substantial increase in over the course of a four-year interval in the proportion of patients dying in the ICU with limitations, an increase in patients discharged alive after treatment withholding decisions, and a reduction in failed resuscitation[15]. During that time mortality remained stable, which provides additional support for our finding that there may not be a stark tradeoff between invasive interventions and survival at the end of life.

Further research is needed to corroborate these findings. If the findings hold, the use of invasive interventions in the ICU for poor prognosis patients who are not responding to ICU care should be fundamentally reconceptualized as a high-risk form of treatment, and decision support tools will be needed to help ICU clinicians determine which patients will likely benefit from invasive interventions, which are not likely to benefit, and which may actually be harmed by them.

Our analysis has several important limitations. This is an observational study and therefore may not imply causality. In addition, it is a single center study, and there may be additional confounders for which we did not adjust. Still, these results raise important questions about current approaches to invasive intervention in patients who are likely to be near the end of life.

**Conclusions**

Despite the expectation that patients would die sooner if invasive interventions were withheld or discontinued, we found no such association. Although it is likely that some individuals died more quickly because endotracheal ventilation and/or vasopressors and inotropes were not administered, for other patients, such invasive interventions may hasten their deaths. Therefore, the assumption that invasive intervention is always life sustaining in this context may be incorrect. For clinicians carrying on end of life discussions with patients and families regarding using, withholding or discontinuing invasive therapies, our findings should provide useful information in terms of the likely impact of such decisions on remaining survival durations.

15. Lesieur, Olivier, et al. "Changes in limitations of life-sustaining treatments over time in a French intensive care unit: A prospective observational study." Journal of critical care (2018).

## Tables

| Independent variables | | | | | | Dependent variables | | | | |
|---|---|---|---|---|---|---|---|---|---|---|
| Year | Number subjects | Age | Gender (male) | SAPS I Score | Elixhauser Score | Vasopressors and inotropes (Yes/No) | Endotracheal Ventilation (Yes/No) | Vasopressors and inotropes (duration in hours) | Endotracheal Ventilation (duration in hours) | Time to death (days) |
| 2002 | 458 | 74 [22] | 50.87% | 22 [9] | 11 [17] | 38.08% | 74.25% | 67.37 [113.12] | 124.87 [176.19] | 5.20 [11.46] |
| 2003 | 490 | 75 [21] | 50.00% | 22 [8] | 12 [18] | 27.00% | 83.75% | 22.50 [74.87] | 112.97 [186.00] | 5.27 [12.53] |
| 2004 | 468 | 72 [24] | 56.83% | 22 [7] | 12 [18] | 32.00% | 85.00% | 32.33 [70.83] | 134.50 [162.87] | 5.04 [10.29] |
| 2005 | 595 | 75 [21] | 49.24% | 22 [8] | 17 [20] | 32.42% | 79.00% | 51.75 [81.25] | 116.00 [155.25] | 5.29 [10.35] |
| 2006 | 555 | 74 [23] | 53.69% | 21 [7] | 17 [19] | 30.92% | 81.50% | 34.00 [68.25] | 111.50 [191.91] | 6.41 [12.12] |
| 2007 | 659 | 75 [22] | 55.38% | 21 [8] | 16 [19] | 26.17% | 59.75% | 21.58 [39.50] | 69.37 [96.57] | 4.58 [11.14] |
| 2008 | 644 | 77 [21] | 51.08% | 21 [8] | 16 [19] | 31.19% | 56.00% | 34.67 [50.88] | 58.50 [116.63] | 4.95 [11.26] |
| 2009 | 666 | 75 [23] | 50.60% | 22 [7] | 19 [19] | 35.03% | 64.00% | 52.27 [86.63] | 80.61 [187.69] | 5.02 [10.98] |
| 2010 | 634 | 75 [21] | 56.94% | 21.5 [7.75] | 17 [20] | 32.19% | 62.17% | 26.83 [52.70] | 59.89 [119.32] | 4.97 [11.07] |

| 2011 | 636 | 75 [22] | 53.14% | 22 [7] | 18 [20] | 33.72% | 65.00% | 38.27 [49.70] | 81.71 [120.23] | 5.54 [12.46] |

Table 1. Characteristics for the cohort of 6,436 patients who died within 30 days from their ICU admission between 2002 and 2011 and trend through the 10 years for all the covariates. For gender, we show the percentage of male for each year. For vasopressors and inotropes (Yes/No), and for endotracheal ventilation (Yes/No) we provide percentage of patients who received the treatments. For all the other covariates we show median values with interquartile ranges.

| Covariate / Cohort | Lower severity score 2002-2005 | Lower severity score 2008-2011 | P-value | Higher severity score 2002-2005 | Higher severity score 2008-2011 | P-value |
|---|---|---|---|---|---|---|
| **Vasopressors and inotropes (duration in hours)** | 41.8 [+/- 86.2] | 34.8 [+/- 82.8] | p-value<0.001 | 24.3 [+/- 77.0] | 26.6 [+/- 68.3] | p-value=0.12 |
| **Endotracheal Ventilation (duration in hours)** | 120.8 [+/- 148.8] | 68.2 [+/- 134.8] | p-value<0.001 | 47.7 [+/- 123.0] | 46.0 [+/- 120.4] | p-value=0.004 |
| **Proportion of Non-Invasive Ventilation Use** | 3.5% | 5.7% | p-value=0.09 | 1.3% | 6.6% | p-value<0.001 |
| **Proportion of Patients with Do-not-resuscitate or** | 14.5% | 25.9% | p-value<0.001 | 14.7% | 20.6% | p-value=0.01 |

| DNR status | | | | | | |
|---|---|---|---|---|---|---|
| Median Time to Death (days) | 7.8 [+/- 8.3] | 8.2 [+/- 8.6] | p-value=0.32 | 2.1 [+/- 6.8] | 2.0 [+/- 6.8] | p-value=0.74 |

Table 2. Comparison of the characteristics of the patients within the lower and higher severity score groups during the earlier years (2002 to 2005) and the later ones (2008 to 2011).

# Figures

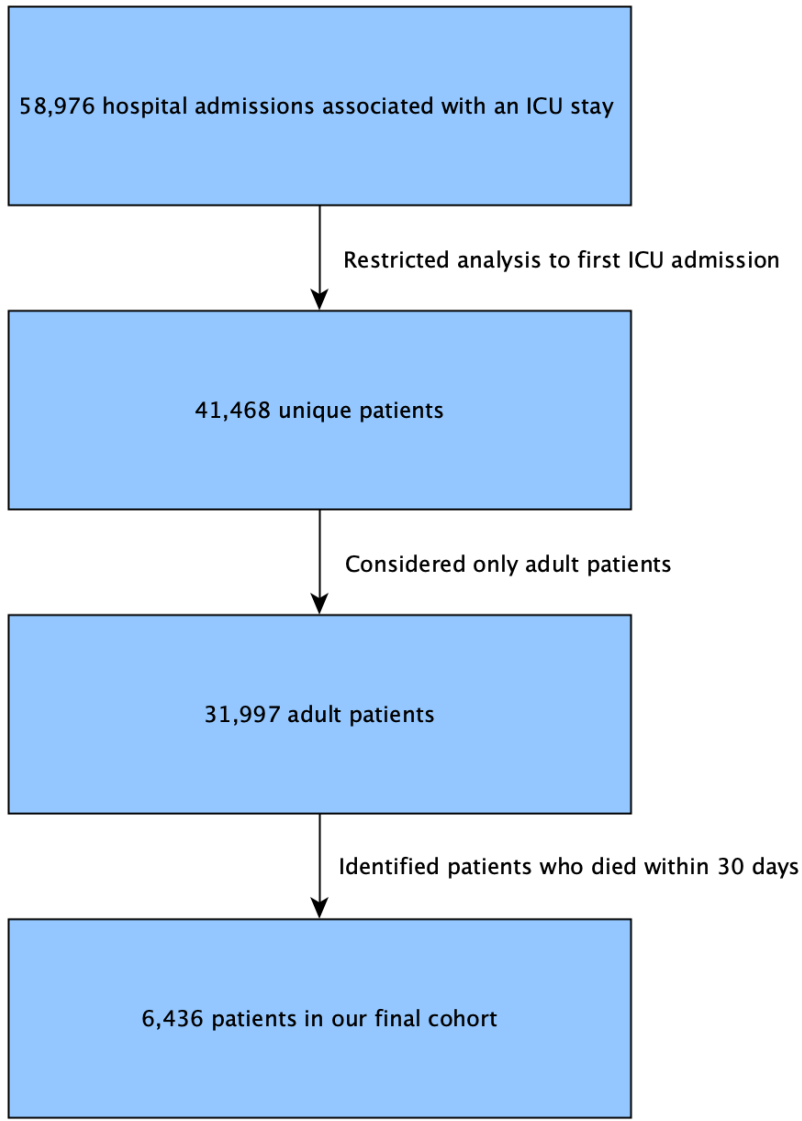

**Figure 1.** Patient Cohort.

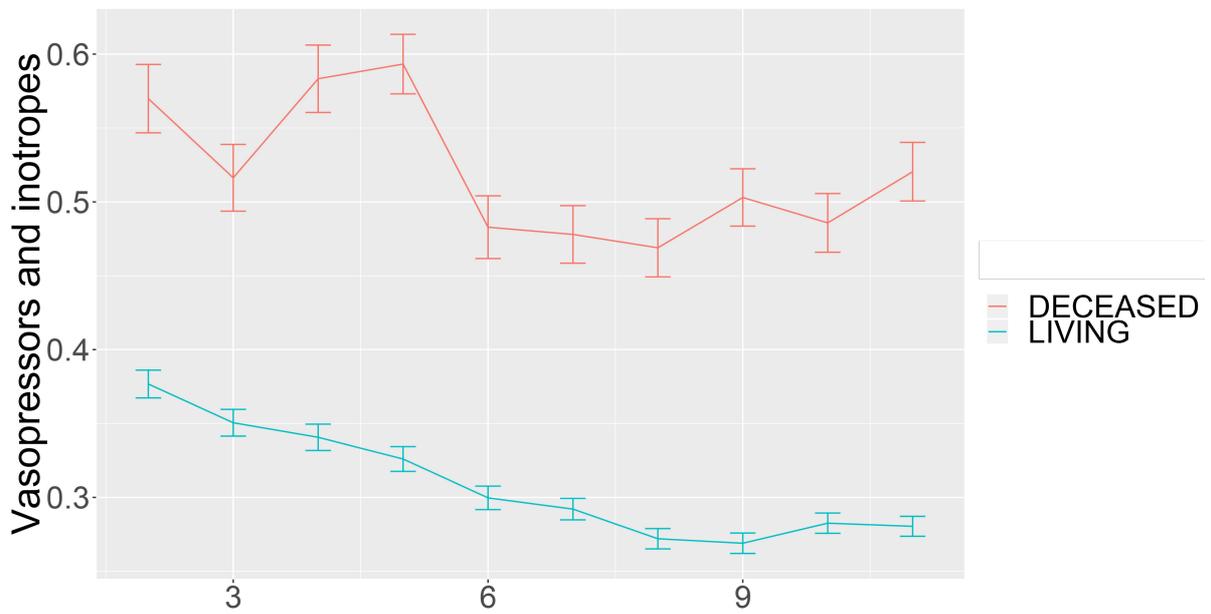

Figure 2. Percentage of patients and standard error who received vasopressors and intropes between 2002 and 2011. Cohort: 6,436 patients who died within 30 days of ICU admission and 40,041 patients who did not die.

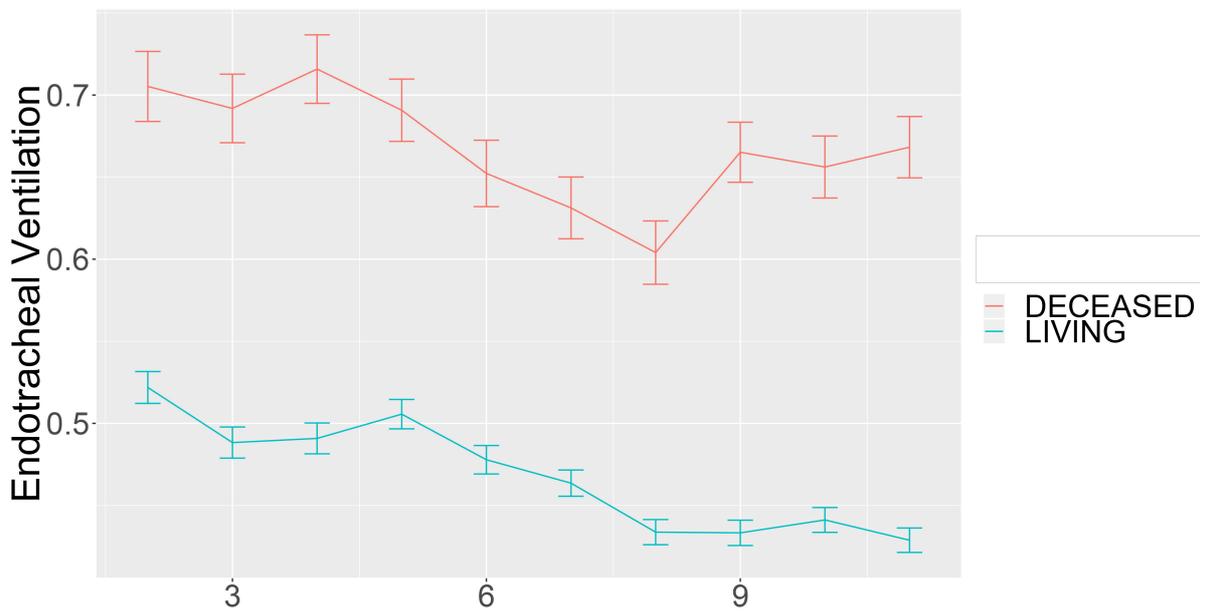

Figure 3. Percentage of patients and standard error who were treated with endotracheal ventilation between 2002 and 2011. Cohort: 6,436 patients who died within 30 days of ICU admission and the 40,041 patients who did not die.